\documentclass{article}

\usepackage{arxiv}

\usepackage[utf8]{inputenc} 
\usepackage[T1]{fontenc}    
\usepackage{hyperref}       
\usepackage{url}            
\usepackage{booktabs}       
\usepackage{amsfonts}       
\usepackage{nicefrac}       
\usepackage{microtype}      
\usepackage{lipsum}
\usepackage{graphicx}
\graphicspath{ {./images/} }

\usepackage{amsmath}

\title{Enhanced rotational diffusion and spontaneous rotation of an active Janus disk in a complex fluid}

\author{
 Marco De Corato \\
  \textit{$^{a}$~Aragon Institute of Engineering Research (I3A), University of Zaragoza, Zaragoza, Spain} \\
  \texttt{mdecorato@unizar.es} \\
   \And
 Paula Mart\'inez-Lera \\
  \textit{$^{a}$~Aragon Institute of Engineering Research (I3A), University of Zaragoza, Zaragoza, Spain} \\
  }

\begin{document}
\maketitle
\begin{abstract}
Active colloids and self-propelled particles moving through microstructured fluids can display different behavior compared to what is observed in simple fluids. As they are driven out of equilibrium in complex fluids they can experience enhanced translational and rotational diffusion as well as instabilities. In this work, we study the deterministic and the Brownian rotational dynamics of an active Janus disk propelling at a constant speed through a complex fluid. The interactions between the Janus disk and the complex fluid are modeled using a fluctuating advection-diffusion equation, which we solve using the finite element method. Motivated by experiments, we focus on the case of a complex fluid comprising molecules that are much smaller than the size of the active disk but much bigger than the solvent. Using numerical simulations, we elucidate the interplay between active motion and fluid microstructure that leads to enhanced rotational diffusion and spontaneous rotation observed in experiments employing Janus colloids in polymer solutions.  
By increasing the propulsion speed of the Janus disk, the simulations predict the onset of a spontaneous rotation and an increase of the rotational diffusion coefficient by orders of magnitude compared to its equilibrium value. These phenomena depend strongly on the number density of the constituents of the complex fluid and their interactions with the two sides of the Janus disk. Given the simplicity of our model, we expect that our findings will apply to a wide range of active systems propelling through complex media.
\end{abstract}

\section{Introduction}

In contrast to passive colloidal particles, active particles harness energy sources or external fields to move through fluidic environments. The obvious examples are living organisms such as bacteria, cells, and small multicellular organisms, that employ different mechanisms to swim. However, in the past 20 years, there has been a large development of synthetic colloids that can self-propel in a way that mimics small living organisms \cite{paxton2004catalytic,howse2007self,ebbens2010pursuit,patino2018fundamental}.  Given their ability to move autonomously and the possibility to functionalize their surface, they represent a promising platform for bioengineering applications such as targeted drug delivery \cite{tang2020enzyme}, microsurgery \cite{li2017micro}, sensing \cite{parmar2018micro}, and cancer treatment \cite{hortelao2018enzyme,simo2024urease}. 

Living organisms and synthetic active colloids often move through complex fluids displaying a mesoscopic scale that is intermediate between their size and the size of the solvent molecules \cite{li2021microswimming,spagnolie2023swimming}. In the case of bacteria, cells, and other small organisms the complex fluid could be any biological fluid, which usually displays many mesoscopic molecules, e.g., biopolymers, proteins, or fibers \cite{spagnolie2015complex}. This is also the case for synthetic active colloids, which are required to move through biological fluids to reach the target of their application \cite{wu2020medical,serra2024catalase}. 

There have been a great number of works focusing on the effect of viscosity, viscoelasticity, and non-Newtonian rheology on the swimming speed of living organisms \cite{lauga2007propulsion,zhu2012self,de2015locomotion,patteson2015running,li2016collective,elfring2016effect,binagia2020swimming,eastham2020axisymmetric,nganguia2020note,arratia2022life} and synthetic active colloids \cite{bechinger2016active,datt2017active,natale2017autophoretic,qiao2020active,sprenger2022active,zhu2024self} 
Comparatively, less is known about the orientational dynamics of these objects in complex fluids. Since the orientation of active particles dictates the direction in which they propel, it is interesting from an engineering perspective to understand what are the phenomena that govern their dynamics. 

In simple fluids, the exchange of momentum between the active particle and the solvent molecules gives rise to a classical rotational Brownian motion, that randomizes its orientation cite{howse2007self}. The Brownian torque reorients the propulsion direction on a timescale inversely proportional to its rotational diffusion coefficient, which decreases as the viscosity of the solvent increases. When suspended in a complex fluid, the additional interactions between the active particle and the mesoscopic constituents can lead to additional contributions to its rotational dynamics. 
At equilibrium, we expect that the microstructure of the complex fluid will slow down the rotation of the active particles. However, unexpected phenomena might arise as the active particles are progressively driven out of equilibrium. The constituents of the complex fluid diffuse slowly and their distribution may be significantly affected by the active motion, which has feedback on the dynamics of the Janus particle.  

Our study is motivated by recent experiments showing that spherical Janus particles moving through a polymer solution experience enhanced rotational diffusion \cite{gomez2016dynamics,lozano2018run,narinder2019active}, and sustained rotation when propelling faster than a critical speed \cite{narinder2018memory,narinder2021work}. 
The enhanced rotational diffusion of active particle has also been observed when propelling through colloidal suspensions of spheres \cite{singh2022interaction}, rods \cite{narinder2022understanding}, and colloidal glasses \cite{lozano2019active}, but the onset of a spontaneous rotation has only been observed in a polymer solution. In addition, the rotational diffusion enhancement is stronger in a polymer solution than in a suspension of colloidal particles.
These differences might be explained by the fact that the colloidal particles are comparable in size to the active probe while polymers are much smaller. 
Such increased coupling between translational and rotational motion of active particles has been observed using numerical simulations \cite{aragones2018diffusion,du2019study,qi2020enhanced,theeyancheri2020translational} in a regime where the constituents of the complex fluid and the active probe are of similar size. 

Here, we investigate how the interplay between active motion and the complex microstructure of the fluid affects the rotational dynamics of active particles. We focus on the regime in which the constituents of the complex fluid are much smaller than the size of the active probe but much bigger than the solvent molecules, which is relevant for a wide class of experimental systems \cite{ignes2022experiments}. Our goal is to elucidate the mechanism that leads to enhanced rotational diffusion and spontaneous rotation observed in experiments.
To do so, we consider a simplified two-dimensional system whereby an active disk propels at a constant speed and interacts with the constituents of the complex fluid through an interaction potential. To isolate the effects of the interactions between the active disk and the complex fluid and further simplify the analysis and we neglect hydrodynamic interactions. 

\section{Governing equations}
\label{sec:eqs}

\begin{figure}[h!]
\centering
\includegraphics[width=9.5cm]{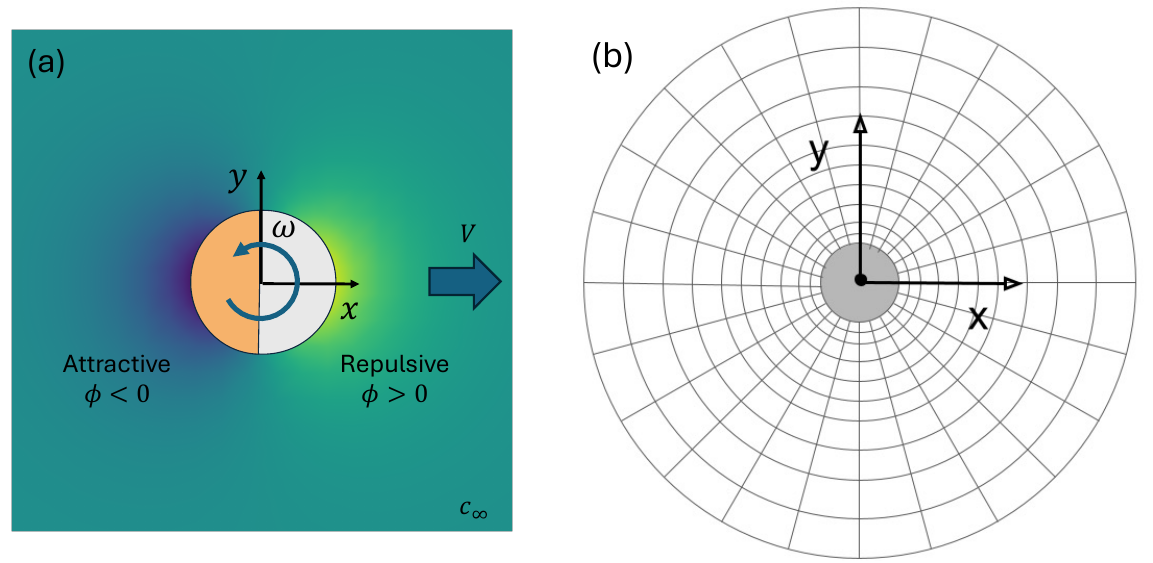}
\caption{(a) Schematic description of an active Janus disk propelling at a constant speed through a complex fluid of number density $c_\infty$. The color scheme represents the interaction potential between the constituents of the complex fluid and the surface of the Janus disk. (b) Schematic example of the computational domain and the structured mesh used in the numerical simulations. The radius of the domain used in the simulations is 50 times the particle radius. The mesh used in the numerical simulations is more refined than that shown here.}
\label{fig:JanusFixed}
\end{figure}

We consider an active Janus disk of radius $R$ propelling through a complex fluid at a constant speed $V$. We employ a co-moving and co-rotating Cartesian coordinate system that is fixed to the center of the active disk and rotates with it (see \ref{fig:JanusFixed}(a)). In this frame of reference, the active Janus disk always propels along the x-axis. We are interested in the case of a complex fluid made of mesoscopic constituents that are much smaller than the size of the active Janus disk but much bigger than the solvent size. The constituents could be nanoparticles, polymers, or other microscopic entities as long as their size is much smaller than $R$. We further assume that the complex fluid is fully described by a scalar field, $c$, which specifies the local number density of its microscopic constituents. By doing so, we are neglecting the internal degrees of freedom of the molecules that constitute the complex fluid, e.g., the end-to-end distance in the case of polymers. We assume that they interact with the Janus disk through a potential energy given by
\begin{equation}\label{eq:potential}
\phi = \phi_0 \,  e^{-\lambda \left|r-R\right|} \cos(\theta) \, \, , 
\end{equation}
where the angle $\theta$ is defined as $\theta =\arccos\left(\frac{x}{r}\right)$, $r$ is the distance from the axes origin $r=\sqrt{x^2+y^2}$, the parameter $\lambda$ is the inverse of the characteristic interaction distance and $\phi_0 $ determines the magnitude of the interaction energy. As shown in Figure \ref{fig:JanusFixed}(a), the interaction is repulsive for $x>0$ and attractive for $x<0$ meaning that the molecules of the complex fluid interact differently with the two sides of the Janus disk and that the disk is propelling towards its repulsive side. This is a reasonable assumption since the two hemispheres of the Janus disk can have different surface chemistry. 

By neglecting intramolecular and hydrodynamic interactions, the number density, $c$, satisfies a stochastic advection-diffusion equation
\begin{equation}\label{eq:diffusionXX}
	\frac{\partial c}{\partial t} - \nabla \cdot \left( D \nabla c  -
	c \mathbf{v} + \frac{D}{k_BT} c \nabla \phi\right)= \nabla \cdot \left( \sqrt{2 D c} \, \boldsymbol{\zeta} \right)\,,
\end{equation}
where $D$ is the diffusion coefficient of the solute, $k_B$ is the Boltzmann constant, $T$ is the temperature and $\mathbf{v}$ is the velocity of the co-moving and co-rotating frame of reference 
\begin{equation}
\mathbf{v}=(-V - \omega r \sin \theta)\boldsymbol{\hat{x}} + \omega r \cos \theta \boldsymbol{\hat{y}} \, \, ,
\end{equation} 
 where $\boldsymbol{\hat{x}} $ and $\boldsymbol{\hat{y}}$ are unitary vectors aligned with the $x$- and, $y$-axes, respectively. The last term on the right-hand side of Eq. \eqref{eq:diffusionXX} represents a fluctuating diffusive flux of solutes determined by the Gaussian random variable, $\boldsymbol{\zeta}$, which has zero mean and variance given by
\begin{equation}
    \langle \boldsymbol{\zeta}(\boldsymbol{x},t)\boldsymbol{\zeta}(\boldsymbol{x}',t')\rangle = \delta(\boldsymbol{x}-\boldsymbol{x}') \delta(t-t') \boldsymbol{I} \, \, . 
\end{equation}
This stochastic term is important when considering the motion of small colloidal particles because the number density of the molecules constituting the complex fluid is small and their thermal fluctuations become relevant. Here, we model the thermal fluctuations using the theory of fluctuating hydrodynamics \cite{de2006hydrodynamic}, which was originally proposed by Landau and Lifschitz \cite{landau2013statistical}. It is based on adding a stochastic forcing term that satisfies the fluctuation-dissipation balance and has been used to study fluctuations in reaction-diffusion equations \cite{kim2017stochastic}, colloidal suspensions \cite{keaveny2014fluctuating,delmotte2015simulating,de2016finite,sprinkle2019brownian}, thin films \cite{sprittles2023rogue}, viscoelastic fluids \cite{hutter2020fluctuating}, and bubble nucleation \cite{gallo2023nanoscale}.

As boundary conditions, we fix the number density $c=c_\infty$ as $r\rightarrow \infty$ and we assume that the disk surface is not penetrable by the solute  
\begin{equation}
    \left( D \nabla c  -
	c \mathbf{v} + \frac{D}{k_BT} c \nabla \phi + \sqrt{2 D c}\boldsymbol{\zeta}\right) \cdot \boldsymbol{n}=0 \, \, ,
\end{equation}
where $\boldsymbol{n}$ is the vector normal to the surface of the active Janus disk. 

To find the angular velocity of the Janus particle we consider the balance of angular momentum 
\begin{equation}\label{eq:torquebal}
\omega=\int_{\Omega_f}{\mathbf{r} \times \frac{c \,  \nabla\phi}{\xi}} d^3\mathbf{x}+\sqrt{\frac{2 \, k_BT}{\xi}}{Z}\,,
\end{equation}
where $\xi$ is the rotational friction coefficient introduced by the friction between the Janus disk and the solvent and $Z$ is a Gaussian white noise that is uncorrelated in time
\begin{equation}
\langle Z(t) Z(t')\rangle =\delta(t-t') \, \, .  
\end{equation}
The first integral on the right-hand side of Eq. \eqref{eq:torquebal} represents the effect of the torque introduced by the interactions between the solute field and the surface of the Janus particle. The second term on the right-hand side represents the effect of the standard Brownian torque introduced by the exchange of momentum with the solvent. 

It becomes apparent from Eq. \eqref{eq:torquebal} that an interaction energy, $\phi$, that depends only on the radial distance, $r$, cannot lead to any angular velocity. In the case of a potential that depends on the radial distance only, its gradient is also directed along the radial coordinate, $\boldsymbol{\nabla} \phi \propto \hat{\boldsymbol{r}}$, where $\hat{\boldsymbol{r}}$ is the unit vector in the radial direction. Therefore, the cross product in Eq. \eqref{eq:torquebal} becomes $\boldsymbol{r} \times \hat{\boldsymbol{r}}=0$ and the first term on the right-hand side vanishes. As a consequence, to have an impact on the rotational dynamics of the Janus disk, the molecules of the complex fluids must experience an interaction that changes with the angle $\theta$. In other words, they need to interact differently with the two sides of the Janus disk (see \eqref{eq:potential}). Here we made the simplest choice of $\phi \propto \cos{(\theta)}$ but other choices would give qualitative similar results to those discussed here.

We remark that our approach based on the stochastic advection-diffusion equation, Eq. \eqref{eq:diffusionXX}, is equivalent to performing Brownian dynamics simulations. This equivalence is explained in Donev and Vanden-Eijnden \cite{donev2014dynamic}, who showed that solving the stochastic partial differential equations, given by Eqs \eqref{eq:diffusionXX}-\eqref{eq:torquebal}, for the number density field $c$ is equivalent to integrating the Brownian dynamics of an independent collection of point-like particles with that interact with the active Janus disk given through the potential given by Eq. \eqref{eq:potential}. The advantage of our approach over Brownian dynamics is the possibility to simulate large domains and millions of diffusing particles at a reduced computational cost. Indeed, most of our simulations are run on a standard desktop computer or a laptop. Using large domains is necessary to avoid finite-size artifacts when the speed of the active Janus disk is large. Finally, care should be taken when integrating Eq. \eqref{eq:diffusionXX} in time. Since the variance of the fluctuating term on the right-hand side is proportional to the number density field, the noise is multiplicative, and different interpretations of the stochastic integral lead to physically different equations, see \cite{hutter1998fluctuation} for a discussion. In this case, the noise in the right-hand side of Eq.  \eqref{eq:diffusionXX} must be interpreted as an Ito stochastic integral \cite{donev2014dynamic}.

By rescaling the distance with $R$ and time using the characteristic diffusion time $R^2/D$ we obtain four dimensionless numbers. 
The dimensionless potential energy $\phi_0^*=\frac{\phi_0}{k_B T}$, the dimensionless inverse interaction length $\lambda^*=\lambda R$, the number of molecules contained in an area proportional to that of the Janus disk $c_{\infty}^*=R^2 c_{\infty}$, the ratio of the disk rotational time to the time required to the solute to diffuse over the disk radius $\xi^*=\frac{ \xi \, D}{R^2 \, k_BT}$ and the P\'{e}clet number based on the translational velocity of the Janus disk $\text{Pe}=\frac{V R}{D}$.

In our work, we explore different values of the dimensionless parameters but we try to connect their value to the value used in recent experiments with active Janus particles in a polymer solution \cite{gomez2016dynamics,narinder2018memory}. We vary the dimensionless number density $c_{\infty}^*$ in the range $c_{\infty}^* \in [500-3000]$, which is close to the value used in experiments. The interaction range of the potential is chosen to be smaller or equal to the radius of the active disk $\lambda^* \in [3-1]$, and its magnitude is chosen to be of the order of a few $k_B T$, $\phi_0^*\in [0.5-2]$. The ratio of the time required for the solute to diffuse over the disk radius to the disk rotational time is not easy to estimate but we expect it to be much larger than one, $\xi^* \gg 1$ because the molecules of the complex fluid are much smaller than the active Janus disk and they diffuse much faster than its rotation time.
Finally, the P\'{e}clet number, $Pe$, represents the main dimensionless number that we explore in our work, since in a typical experiment the speed of the active particle can be controlled. Since the diffusion coefficient of polymers is typically of the order of $D\approx 10^{-11}-10^{-12} \rm{m}/\rm{s^2}$, the P\'{e}clet number can be easily of order one for micrometer-sized active particles propelling at $V \approx 1 \rm{\mu m}/\rm{s}$.

\section{Numerical method}
To investigate the dynamics of the active particle, we solve Eqs. \eqref{eq:diffusionXX}-\eqref{eq:torquebal} numerically in a two-dimensional domain comprised of two concentric disks. In Figure \ref{fig:JanusFixed}(b), we report a schematic example of the computational domain and the mesh used in the numerical simulations. To prevent finite-size effects, the far-field boundary condition is applied at the edge of the outer disk, which is placed at a distance of $50R$ from the origin. We use a structured quadrilateral mesh whereby the size of the elements follows a geometric progression in the radial direction towards the external boundary, with finer elements closer to the particle. In this region, we expect larger solute gradients due to the interplay of advection and the interaction between the solute and the active Janus disk. The mesh has $N_r=50$ elements in the radial direction and $N_{\theta}=48$ elements discretizing the circumference, which we checked to yield numerically convergent results.
We discretize the solution in space using a continuous finite element Galerkin discretization with linear shape functions defined on the quadrilateral elements (See Figure \ref{fig:JanusFixed}(b)). To ensure that the discretized equations satisfy the fluctuation-dissipation balance, the fluctuating flux in the right-hand side of Eq. \eqref{eq:diffusionXX} deserves special treatment. To do so, we follow our recent finite element implementation of stochastic diffusion equations \cite{martinez2024finite}. Nevertheless, in the appendix \ref{sec:numericalmethod}, we include a detailed derivation of the numerical method employed.

\section{Results}
\label{sec:results}
To study the dynamics of the active Janus disk in detail, we first focus on its behavior in the absence of thermal fluctuations. We then proceed to include the thermal fluctuations in Sections \ref{sec:equilibrium_brownian} and \ref{sec:nonequilibrium_brownian} to characterize its rotational Brownian dynamics.

\subsection{Spontaneous rotation of the active Janus disk}\label{sec:sustainedrotation}
In the absence of a time-dependent stochastic term, Equation \eqref{eq:diffusionXX} has a steady solution that satisfies the steady advection-diffusion equation
\begin{equation}\label{eq:diffusionSteady}
	 \nabla \cdot \left( D \nabla c  -
	c \mathbf{v} + D c \nabla \phi\right)= 0\,.
\end{equation}
with boundary conditions $c \rightarrow c_\infty$ as $r\rightarrow \infty$, zero flux on the boundary of the Janus disk
\begin{equation}
\left( D \nabla c  -
	c \mathbf{v} + D c \nabla \phi\right) \cdot \boldsymbol{n}= 0 \, \, ,
\end{equation}
and the angular velocity of the active Janus disk given by
\begin{equation}
\omega=\int_{\Omega_f}{\mathbf{r} \times \frac{c \,  \nabla\phi}{\xi}} d^3\mathbf{x}\,.
\end{equation}
The resulting steady-state equations are nonlinear because the rotational velocity, $\omega$, that appears in the advection-diffusion equation, given by Eq. \eqref{eq:diffusionSteady}, depends on the number density field via Eq. \eqref{eq:torquebal}. To solve these equations numerically, we use the Newton-Raphson method.  
\begin{figure}[h!]
    \centering
    \includegraphics[width=9.5cm]{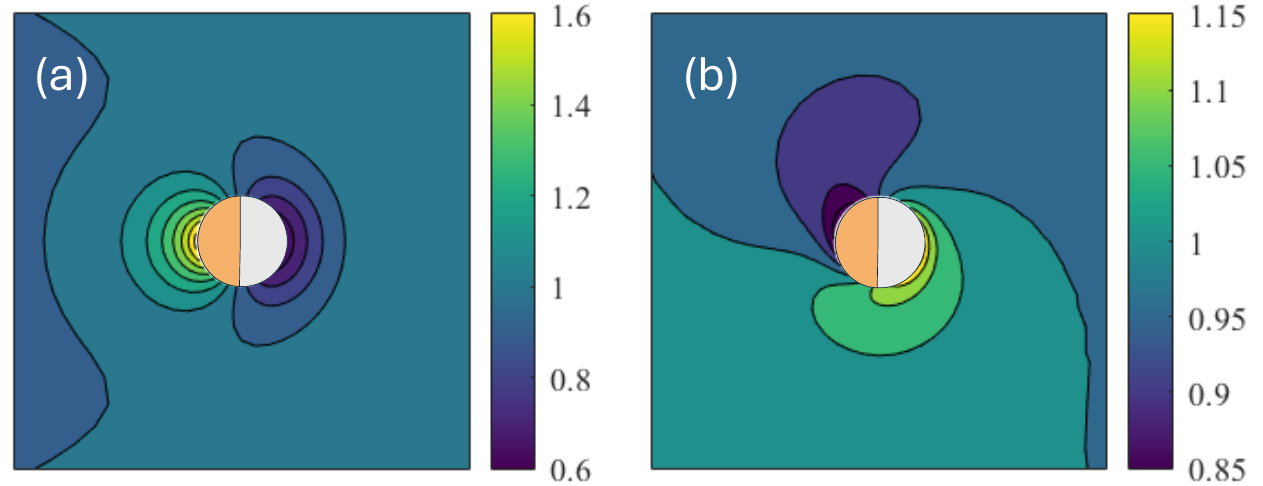}
    \caption{Contour plots of the normalized number density field, $c^*/c^*_{\infty}$, for the case of $c^*_{\infty}=1000$, $\phi_0^*=1$, $\lambda^*=1$, $\xi^*=250$. (a) Symmetric solution with zero angular velocity ($\text{Pe}=0.5$) and (b) Symmetry with respect to the $x-$axis is broken and the active Janus disk rotates ($\text{Pe}=1.5$).}
    \label{fig:cfield}
\end{figure}

The problem of an active Janus disk translating at a constant speed is symmetric with respect to the $x$-axis (See Figure \ref{fig:JanusFixed}(a)), we thus expect the zero angular velocity to be zero. However, previous works on chemically-active particles, microswimmers, and active droplets have shown that the symmetry can break and the particle can spontaneously rotate \cite{de2021spontaneous,morozov2019orientational}. 
Figure \ref{fig:cfield} shows two examples of the computed solute fields using the Newton-Raphson method. In Figure \ref{fig:cfield}(a) we fix $\text{Pe}=0.5$ and we find that the particle does not rotate. The solute field is symmetric with respect to the $x$-axis and displays the characteristic wake due to the advection due to the reference frame moving with the active Janus disk, which was also observed in recent simulations \cite{dai2023hydrodynamic}. Instead in Figure \ref{fig:cfield}(b), obtained for $\text{Pe}=1.5$, the particle displays a nonzero angular velocity, $\omega$, and the solute field has lost the symmetry around the $x$-axis in the moving and rotating frame of reference. 

\begin{figure}[h!]
\centering
\includegraphics[width=8.0cm]{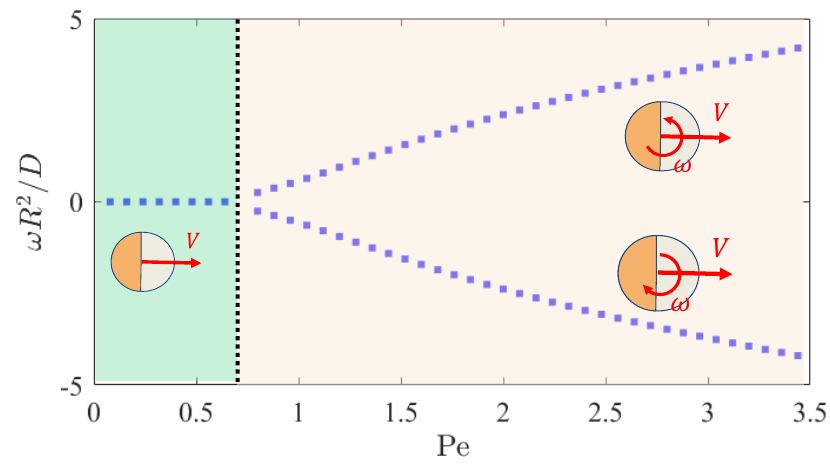}
\caption{Dimensionless angular velocity as a function of the P\'{e}clet number, with $[c^*_{\infty}, \phi_0^*, \xi^*, \lambda^*]=[1000, 1, 250,1]$.}
\label{fig:omega_vs_Pe}
\end{figure}

By repeating the simulations for different values of $Pe$, we can construct a plot of the dimensionless angular velocity as a function of $Pe$. This is reported in Figure \ref{fig:omega_vs_Pe} by fixing the remaining parameters $[c^*_{\infty}, \phi_0^*, \xi^*, \lambda^*]=[1000, 1, 250,1]$
Below a critical value of $\text{Pe}$, the active Janus disk does not rotate and $\omega=0$. Above the critical value of $\text{Pe}$, however, the Janus disk starts to rotate with an angular speed that increases with increasing translational velocity. Since the Janus disk can rotate in either direction, the curve presents two different branches, one displaying positive $\omega$ and one displaying negative $\omega$. The instability mechanism found for the active Janus disk is analogous to that found for chemically-active particles, microswimmers, and active droplets \cite{de2021spontaneous,morozov2019orientational}. Here the advection of the solute comprising the complex fluid by the translating Janus disk accumulates solute on its repulsive side, which ultimately becomes unstable to rotational perturbations. 

\begin{figure}[h!]
\centering
\includegraphics[width=6cm]{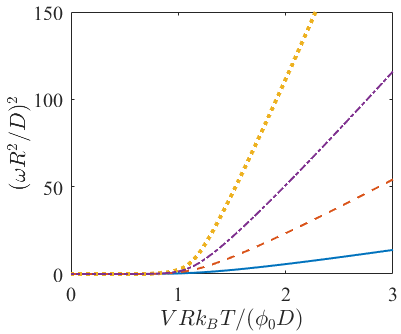}
\caption{Square of the computed angular velocity as a function of the ratio $V R/\phi_0 D$, with $[c^*_{\infty}, \phi_0^*, \xi^*, \lambda^*]=[1000, 1, 250,1]$ (solid), $[1000, 1, 80,1]$ (dashed), $[1000, 2, 80,1]$ (dotted), $[2000, 1, 80,1]$ (dash-dotted).}
\label{fig:Sqomega_vs_Pe}
\end{figure}

The critical P\'{e}clet number changes with the dimensionless numbers considered in the simulation. However, we can estimate its value number by balancing the advective flux, which scales with $V c$, and the flux due to the potential interaction which scales as $D \, c \, \phi_0/R \, k_BT$. By balancing the two fluxes we estimate the critical Péclet number as $\text{Pe}_c \approx \phi_0^*$ or, in dimensional form, a critical velocity $ V_c \approx \phi_0 D/ k_BT \, R$. 
This scaling is demonstrated in Figure \ref{fig:Sqomega_vs_Pe}, where we plot the squared value of the normalized angular velocity as a function of the ratio $V R/\phi_0 D$. Above $V\approx \phi_0 D/k_BT R$, the squared value of $\omega$ is approximately linear with the ratio $V R k_BT/\phi_0 D$. We note that this estimate only works if advection and the potential interaction are the dominant mechanisms and diffusion can be neglected.

\begin{figure}[h!]
\centering
\includegraphics[width=8.5cm]{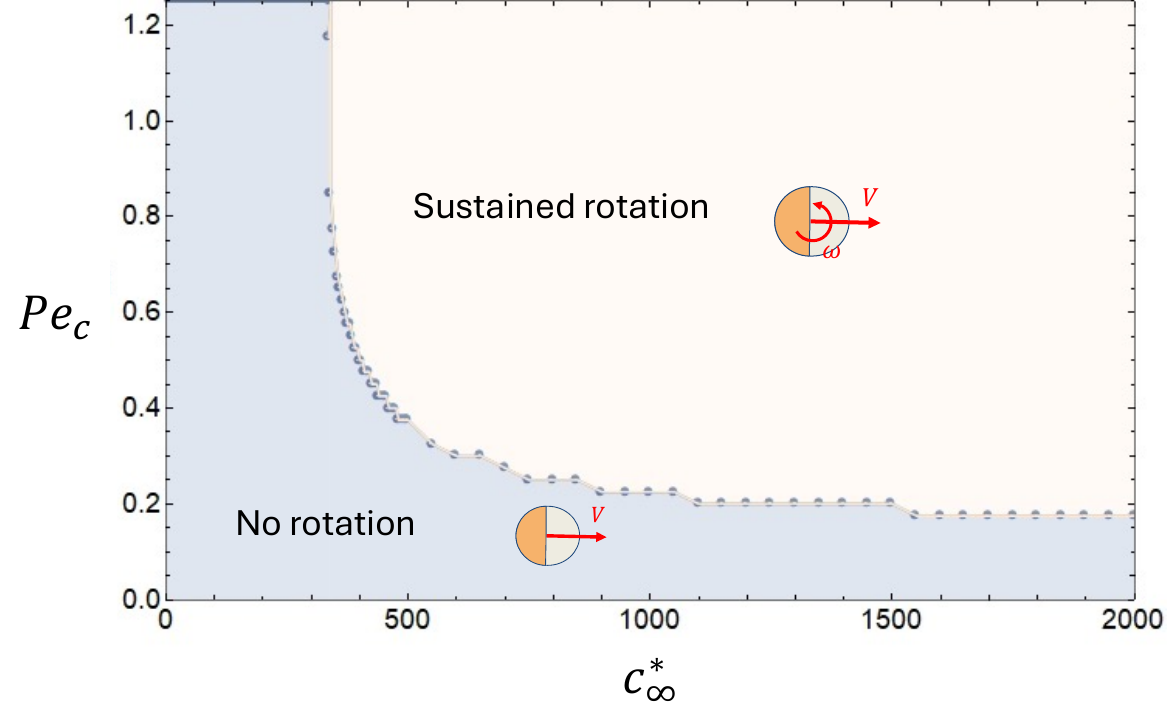}
\caption{Critical P\'{e}clet of active Janus disk that separates regimes of no rotation from regimes of spontaneous rotation as a function of the dimensionless number density of the complex fluid. The remaining parameters are $[ \phi_0^*, \xi^*, \lambda^*]=[1, 100, 3]$.}
\label{fig:solution_map}
\end{figure}

In Figure \ref{fig:solution_map}, we report a plot of the critical P\'{e}clet number as a function of the dimensionless number density of the complex fluid, $c_\infty^*$. As expected, the critical P\'{e}clet number departs from the estimate $\text{Pe}_c \approx \phi_0^*$ for small values of $c_\infty^*$. In this region the diffusive flux becomes progressively more relevant and $Pe_c$ diverges at a given $c_\infty^*$. No spontaneous rotation is possible for systems with a number density $c_\infty^*$ smaller than a critical value, which depends on the system parameters. This could be the reason why this instability has not been observed in previous simulations \cite{qi2020enhanced, theeyancheri2020translational}. In these works, the authors considered active particles comparable in size with the molecules of the complex fluid, which limits $c_\infty^*$ to small values, possibly smaller than the critical value required to observe the spontaneous rotation instability. This could also explain why the spontaneous rotation instability was observed experimentally for active particles propelling through a polymer solution where the polymers are small and their number density is large \cite{narinder2018memory} but not in the case of a colloidal suspension \cite{lozano2019active,abaurrea2020autonomously,narinder2022understanding} where the constituents of the complex fluid are large and comparable to the size of the active particle.
We expect that the onset of the instability leading to the spontaneous rotation significantly influences the Brownian rotations of the active Janus disk.

\subsection{Rotational Brownian dynamics of the Janus disk at equilibrium}\label{sec:equilibrium_brownian}
\begin{figure*}[h!]
\centering
(a)\includegraphics[height=4.9cm]{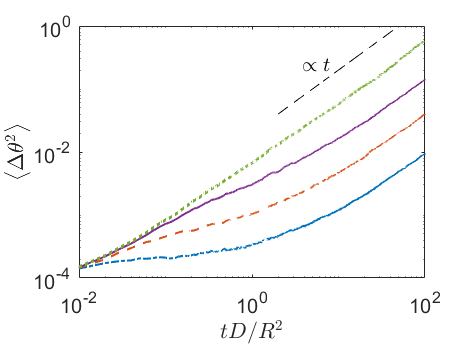}
(b)\includegraphics[height=4.7cm]{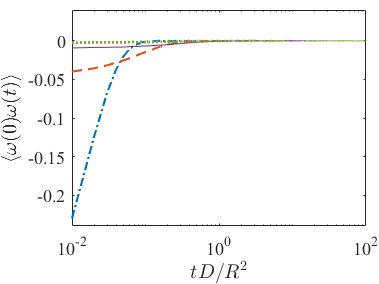}
\caption{(a) Mean squared angular displacement at equilibrium as a function of dimensionless time. (b) Autocorrelation of the angular velocity at equilibrium plotted as a function of dimensionless time. The dimensionless parameters used are $[\phi_0*,c_\infty*,\xi,\lambda]=[2,1000,250,1]$ (dash-dotted), [1,1000,250,1] (dashed), [0.5,1000,250,1] (solid), [0.5,1000,250,3] (dotted)}
\label{fig:MSAD_equil}
\end{figure*}
To investigate the Brownian reorientation of the active Janus disk, we run unsteady simulations of the stochastic advection-diffusion equation \eqref{eq:diffusionXX} with thermal fluctuations. We begin by analyzing the rotational Brownian dynamics of the active Janus disk at equilibrium $V=0$. In all of our calculations, we used a timestep size $\Delta t=0.01 R^2/D$, which we checked to give convergent results. To compute ensemble-averaged quantities for each set of dimensionless numbers, we simulate $10^6$ timesteps, which is equivalent to $10^4 R^2/D$ characteristic times. We begin our simulations from the equilibrium distribution and we discard the first $10^3$ steps to avoid effects due to the initial conditions. At each time step, we compute the angular velocity of the active Janus disk and we calculate the angular displacement, $\Delta \theta$, with respect to a fixed frame as
\begin{equation}
\Delta \theta (t) = \int_0^t \omega(\tau) \, d\tau \, \, .
\end{equation}

In Figure \ref{fig:MSAD_equil}(a), we report the mean squared angular displacement (MSAD), defined as $\langle \Delta \theta (t)^2 \rangle$, for different values of the dimensionless parameters. At long times all the curves show a diffusive regime with a linear MSAD increase as a function of time. As the strength or the range of the interaction potential and the number density increase, we observe the appearance of a transient subdiffusive regime at short times and a linear MSAD at longer times.

To better understand these results, we calculate the autocorrelation of the angular velocity, defined as $\langle \omega(0) \omega(t) \rangle$, for the same value of the dimensionless parameters used in Figure \ref{fig:MSAD_equil}(a). The autocorrelation of the angular velocity is plotted in Figure \ref{fig:MSAD_equil}(b). As the strength or the range of the interaction potential and the number density increase, an increasing negative correlation appears, which denotes an increasingly hindered rotational motion of the Janus disk. The negative autocorrelation then decays to zero exponentially over a timescale comparable to $R^2/D$. The onset of a negative correlation is a consequence of the interaction between the constituents of the complex fluid and the Janus disk. The rotations of the Janus disk are hindered because they change the number density compared to its equilibrium value. The density perturbations tend to be restored to their equilibrium value by a rotation in the opposite direction, which explains the negative autocorrelation of the angular velocity. This phenomenon is analogous to the hindered translational diffusion of charged particles that interact with the surrounding cloud of ions \cite{gorti1984determination,medina1985electrolyte}, with the difference that the rotational diffusion is slowed down in the case of the Janus disk. 
\begin{figure}
\centering
\includegraphics[width=9.0cm]{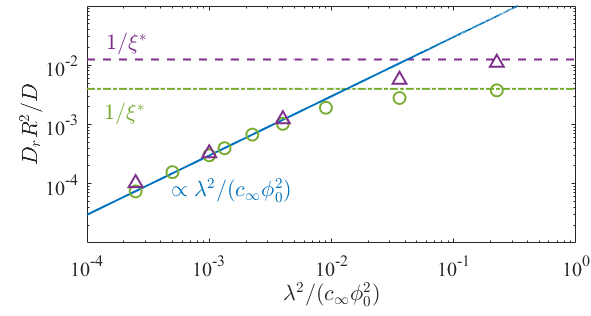}
\caption{Normalized rotational diffusion coefficient for $V=0$ with $\xi^*=80$ ($\triangle$) and $\xi^*=250$ ($\circ$). Computed with different values for parameters $\phi_0^*$ ($0.5$, $0.8$, $1$ or $2$), $\lambda^*$ ($1$ or $3$) and $c^*_{\infty}$ ($750$, $1000$ or $2000$). }
\label{fig:DRnorm0}
\end{figure}
To confirm this, Figure \ref{fig:MSAD_equil}(b) shows that the correlation becomes progressively more negative as the concentration grows or the interaction potential becomes stronger. The increased negative correlation observed in the simulations explains the slowdown of the rotational diffusion and the onset of the subdiffusive regime shown in Figure \ref{fig:MSAD_equil}(a). Specifically, the MSAD is subdiffusive as long as a negative correlation between angular velocities persists and then becomes diffusive when the autocorrelation finally decays to zero. 
\begin{figure*}[h!]
\centering
\includegraphics[width=5.0cm]{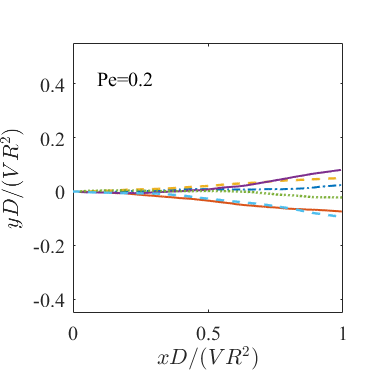}
\includegraphics[width=5.0cm]{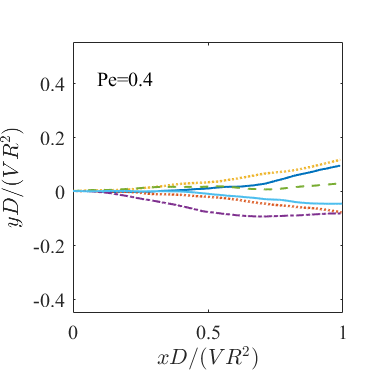}
\includegraphics[width=5.0cm]{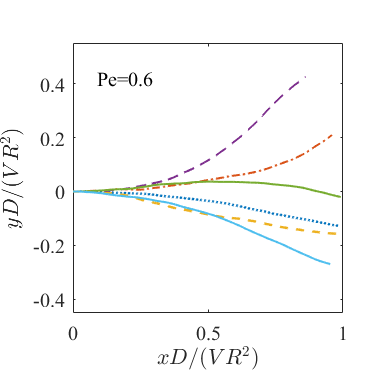}
\includegraphics[width=5.0cm]{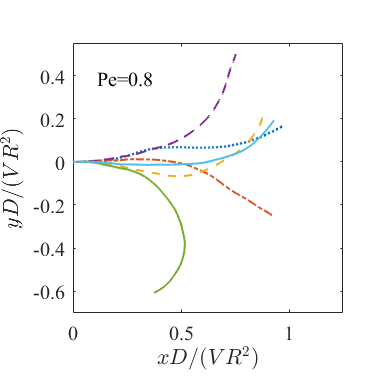}
\includegraphics[width=5.0cm]{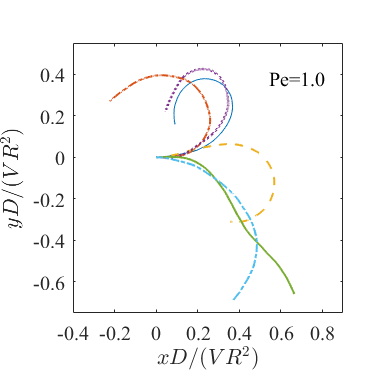}
\includegraphics[width=5.0cm]{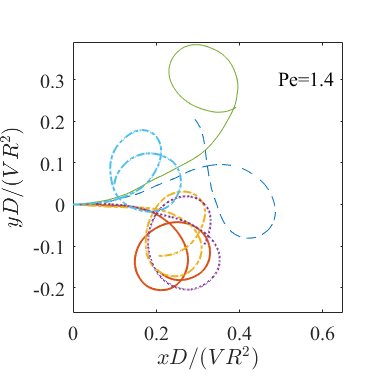}
\caption{Sample trajectories of active Janus disks in a fixed reference frame plotted for $Pe=0.2$, $0.4$, $0.6$, $0.8$, $1.0$ and $1.4$. The Janus disks start at the origin of the axes ($[0,0]$) with an initial particle orientation along $x-$axis. Each line represents a different trajectory that lasts a total time equal to $10 R^2/D$. The remaining dimensionless parameters are given by  $\phi_0*=1$, $\lambda^*=1$, $c^*_{\infty}=1000$, $\xi^*=250.$}
\label{fig:trajectories}
\end{figure*}

We can use the autocorrelation of the angular velocity to calculate the rotational diffusion coefficient of the Janus disk using the Green-Kubo relation
\begin{equation}\label{eq:greenkubo}
    D_r = \int_0^{\infty}{\left<\omega(0) \omega(t)\right>} dt\,.
\end{equation}
In Figure \ref{fig:DRnorm0}, we find that the dimensionless rotational diffusion coefficient, $D_r R^2/D$, calculated for different values of dimensionless parameters can be divided in two regimes depending on the dimensionless number 
\begin{equation}
    \Lambda_2 \propto \frac{\lambda^{*^2}}{\phi_0^{*^2} c_{\infty}^*} \,.
\end{equation}

This number quantifies the strength of the coupling between the fluctuations of the number density of the complex fluid and the angular velocity of the Janus disk.
For large values of $\Lambda_2 $ the interaction between the Janus disk and the number density field is weak and the diffusion coefficient is given by the friction with the solvent $D_r \approx k_BT/\xi$ or, in dimensionless form,  $D_rR^2/D \approx 1/\xi^*$ (see Figure \ref{fig:DRnorm0}). Conversely, Figure \ref{fig:DRnorm0} shows that, for small values of $\Lambda_2 $, the coupling between the Janus disk and the number density field is strong and the rotational diffusion coefficient is always smaller than $k_BT/\xi$ and it scales as $D_r \propto \Lambda_2$. In this regime, the constituents of the complex fluid strongly hinder the rotational diffusion of the Janus disk.

\subsection{Rotational Brownian dynamics of the Janus disk out of equilibrium}\label{sec:nonequilibrium_brownian}
We now turn to the case of Janus disks driven out of equilibrium by a constant nonzero translational velocity $V$. To calculate ensemble average quantities in this case, we perform between $2 \, \cdot 10^3$ and $10^4$ independent simulations, each starting from the solution of the steady-state equation, given by Eq. \eqref{eq:diffusionSteady}, obtained with the Newton-Raphson method. Each independent simulation is integrated in time for 50 or 100 dimensionless times (made dimensionless using  $R^2/D$) with a timestep size $\Delta t= 10^{-2} R^2/D$. To compute the results, we discard the first 250 timesteps to let the system thermalize.  
\begin{figure*}
\centering
(a)\includegraphics[width=8cm]{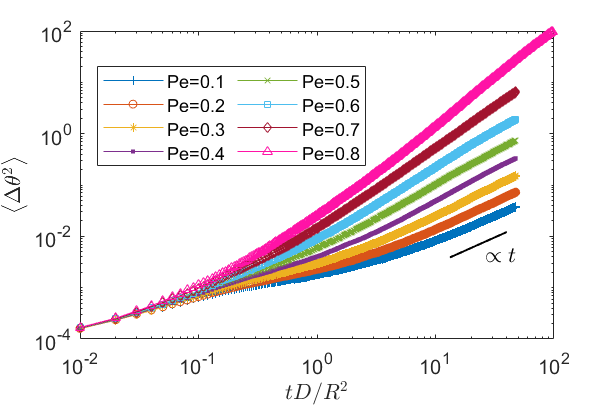}
(b)\includegraphics[width=8cm]{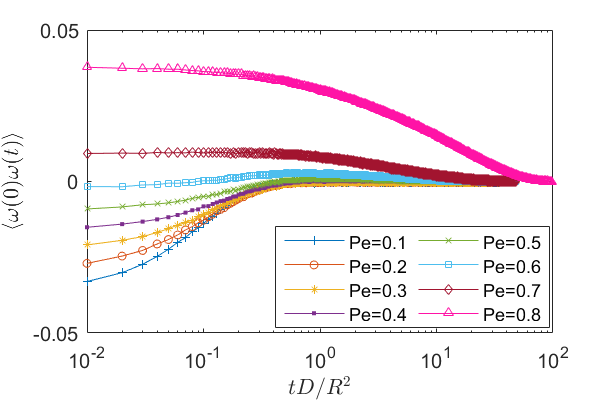}
\caption{(a) MSAD as a function of the dimensionless time and (b) autocorrelation of the angular velocity as a function of the dimensionless time plotted for different values of $Pe$. The remaining dimensionless numbers are $\xi^*=250$, $c_\infty^*=1000$, $\phi^*=1$, and $\lambda^*=1$. }
\label{fig:msadnoneq}
\end{figure*}

In Figure \ref{fig:trajectories}, we report the trajectories of an ensemble of Janus disks in a fixed reference frame at increasing $V$ below and above the critical P\'{e}clet number. As the velocity increases, the trajectories become progressively more curved, which is a signature of an increased rotational diffusion that changes the direction of motion of the active Janus disk. As the particle approaches and surpasses the critical velocity, it increasingly describes circular trajectories, which is a signature of a nonzero angular velocity. In this regime, the Janus disk can switch between clockwise and anticlockwise rotations, suggesting that the angular velocity is not constant but can jump between positive and negative values.

To confirm these findings, we compute the MSAD for different values of the dimensionless numbers. In Figure \ref{fig:msadnoneq}(a), we report the MSAD as a function of time for $c_\infty^*=1000$, $\xi^*=250$, $\phi_0^*=1$. For this set of parameters, the critical P\'{e}clet number obtained from the steady-state solution neglecting the fluctuations is $Pe_c \approx 0.8$. All the curves display a linear behavior at very short times, which is the signature of the angular Brownian motion due to the interactions with the solvent. Then, the MSAD transitions to a regime that is dominated by the interactions with the molecules of the complex fluid. For small values of the Peclet number the long-time MSAD is a linear function of time. However, as the P\'{e}clet number increases the MSAD becomes progressively more superdiffusive. 
Results obtained for different values of the dimensionless parameters show qualitatively similar behavior.

To better understand the behavior of the MSAD, in Figure \ref{fig:msadnoneq}(b) we plot the autocorrelation of the angular velocity as a function of $Pe$ for the same dimensionless parameters used in Figure \ref{fig:msadnoneq}(a). We observe two trends as the $Pe$ is progressively increased and approaches $Pe_c$. First, the negative correlation of the angular velocity observed for small values of $Pe$ switches to a positive correlation as $Pe$ is increased. Second, the correlation of the angular velocity persists over a longer timescale, showing a progressively slower decay as $Pe$ increases.  
This behavior of the autocorrelation is consistent with the superdiffusive behavior of the MSAD observed in Figure \ref{fig:msadnoneq}(a). 
\begin{figure*}[h!]
\centering
\includegraphics[width=5.5cm]{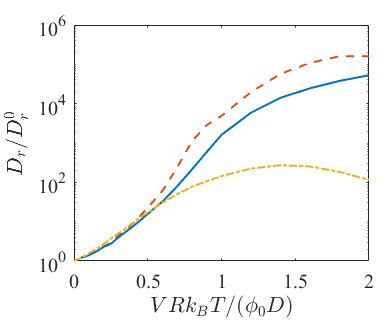}
\includegraphics[width=5.5cm]{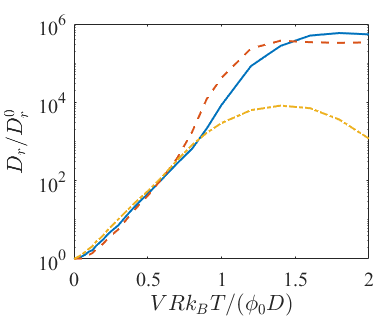}
\caption{Normalized rotational diffusion coefficient for an active disk driven out of equilibrium. (left) $\phi_0=1$ and (right) $\phi_0=2$. The remaining parameters are $[c^*_{\infty}, \xi^*, \lambda]=$  $[1000, 250, 1]$ (solid), $[1000, 80, 1]$ (dashed), $[1000,250, 3]$ (dash-dot).}
\label{fig:DRnorm_vs_V}
\end{figure*}

We can use the Green-Kubo relation, given by Eq. \eqref{eq:greenkubo}, to calculate the rotational diffusion coefficient of the active Janus disk as it is driven out of equilibrium.
Figure \ref{fig:DRnorm_vs_V} shows results for the value of the normalized rotational diffusion coefficient as a function of the translational velocity $V$ of the particle, for different values of the parameters. The rotational diffusion coefficient is normalized with its value at equilibrium $D_r^0$. 
The curves show two distinct regimes, depending on whether or not spontaneous rotation dominates the behavior of the particle. For small values of the velocity ($V<V_c$), the rotational diffusion coefficient increases exponentially with the velocity, reaching a saturation point around the critical velocity, above which $D_r$ shows saturation as $V$ keeps increasing.

For subcritical velocities, the different curves for $D_r/D_r^0$ as a function of the ratio $V R k_B T/\phi_0 D$ collapse on a single curve for a given $\phi_0^*$, independently of the value of $c_{\infty}^*$, $\lambda^*$ and $\xi^*$. As the potential decreases, diffusion effects on the constituents number density become comparatively more important and produce some damping on the rotational response of the particle, as evidenced by the fact that the slope of the curves for $V \ll V_c$ decreases with $\phi_0^*$, and that, for a lower value of $\phi_0^*$, the curves of $D_r^*$ start to deviate from each other for a lower value of $V R k_B T/\phi_0 D$.

\section{Discussion and conclusions}
In this work, we studied the rotational dynamics of an active Janus disk propelling through a complex fluid at a constant speed. We assumed that the complex fluid can be characterized by the local number density of its constituents, which are much smaller than the Janus disk. Because of this, our system is different from that considered in \cite{aragones2018diffusion,qi2020enhanced, theeyancheri2020translational,abaurrea2020autonomously,chepizhko2020random,narinder2022understanding}, who studied the case in which the complex fluid was made of constituents of size comparable or larger than that of the active particle. Similarly to what was done in \cite{de2021spontaneous}, we assumed that the molecules of the complex fluid interact differently with the two sides of the Janus disk. We remark that this choice of the interaction potential is necessary to observe the effects of the complex fluid on the rotational dynamics of the Janus disk. Spatial variations of the number density from their equilibrium value cannot lead to torques if the interaction depends on the radial coordinate only. In this case, the rotational dynamics are unchanged compared to the case of pure solvent. 
We include the thermal fluctuations of the molecules constituting the complex fluid through a fluctuating diffusive flux and we use the finite element method to solve the resulting stochastic advection-diffusion equation governing the number density of the complex fluid. 

We first study the deterministic behavior of the Janus disk by neglecting the thermal fluctuations. We find that, as the propulsion velocity of the Janus disk increases beyond a critical velocity, it spontaneously starts to rotate. The transition is governed by an instability that is similar to what is found for microswimmers, chemically active particles\cite{de2021spontaneous}, and active nematic droplets \cite{morozov2019orientational}. The similarity between these different models suggests that the instability is a generic feature and doesn't depend on a specific model of the active particle. It appears that any mechanism that accumulates solute on a repulsive side and depletes from an attractive side of the active particle eventually leads to a spontaneous rotation. The critical P\'eclet number above which the Janus disk spontaneously rotates depends on the ratio of advective and potential fluxes, provided that the number density of constituents of the complex fluid is above a given threshold. We find that no spontaneous rotation is possible if the density of the constituents of the complex fluid is smaller than a critical value. This could explain why this instability has not been observed in simulations considering active particles comparable in size to the constituents of the complex fluid \cite{aragones2018diffusion,qi2020enhanced, theeyancheri2020translational}. 

We then included thermal fluctuations and studied the rotational Brownian dynamics of the active Janus disk in the complex fluid both at equilibrium and out of equilibrium. At equilibrium, we find that the interaction with the constituents of the complex fluid slows down the rotational diffusion compared to the case of pure solvent. The rotational diffusion decreases as the concentration of the complex fluid increases or as the interaction potential becomes stronger or longer ranged. The mechanism responsible for the hindered rotational diffusion is similar to that responsible for the hindered translational diffusion of charged colloids suspended in an ionic solution \cite{gorti1984determination,medina1985electrolyte}.  As the active Janus disk is driven out of equilibrium by its propulsion, its rotational velocity becomes positively correlated over a long time, and its rotational diffusion increases by orders of magnitudes compared to its equilibrium value. As a result, the trajectories traced by the active disk become progressively more curved and become circular as the velocity exceeds the critical velocity whereby it spontaneously rotates. The enhancement of the rotational diffusion of the active disk due to its propulsion in a complex fluid appears to be much stronger than the enhancement of the translational diffusion of active probes \cite{zia2010single,hoh2016force} and active particles \cite{rizkallah2022microscopic}.

Despite being simplified and two-dimensional, the model presented here displays qualitative features similar to those observed in the experiments of Gomez-Solano et al \cite{gomez2016dynamics} and Narinder et al. \cite{narinder2018memory,narinder2021work}. They studied active Janus colloids driven in a polymer solution by a phase-separating solvent\cite{buttinoni2012active,gomez2020transient,decayeux2021spontaneous,decayeux2022conditions}. First, as the propulsion velocity of the active Janus disk increases, the rotational diffusion of the active disk increases exponentially displaying progressively more curved trajectories. The rotational diffusion coefficient can increase by several orders of magnitude compared to its equilibrium value before reaching a plateau for velocities larger than the critical velocity. Second, a spontaneous rotation is observed beyond a critical velocity. Nevertheless, we should note that the experimental system used by Gomez-Solano et al \cite{gomez2016dynamics}  and Narinder et al. \cite{narinder2018memory} is much more complicated compared to the simple model presented here. Therefore, we cannot exclude that other mechanisms that are not included in our model such as polymer elasticity, stress relaxation, local solvent phase separation, and wall confinement play a role in the experimental observations. The effect of the solvent demixing into polar and nonpolar phases on the polymer distribution could be studied using the framework developed in a recent paper \cite{de2022theory}. 

We emphasize that our findings depend strongly on the direction of motion of the active Janus disk. In our simulations, we have considered a disk that propels in the direction of the repulsive side. However, if the disk propels along the attractive side instead, the disk does not display enhanced rotational diffusion or the spontaneous rotation instability. The sensibility of our findings on the direction of motion could explain why they have been observed in some systems \cite{gomez2016dynamics,narinder2018memory} but not in others \cite{saad2019diffusiophoresis}.

To conclude, our results help elucidate the interplay between active motion, potential interactions, and concentration fluctuations that lead to the enhanced rotational Brownian dynamics and the spontaneous of active particles in a complex fluid.

\bibliography{biblioFH_abbrev.bib} 
\bibliographystyle{ieeetr}

\appendix

\section{Numerical method for the stochastic advection-diffusion equation}
\label{sec:numericalmethod}
To solve the stochastic advection-diffusion equation
given by Equation \eqref{eq:diffusionXX}
we first write a weak formulation of the equation by multiplying by the complex conjugate of a test function $q$ and integrating in the two-dimensional domain, to obtain
\begin{equation}
	\int{q^* \frac{\partial c}{\partial t}}d^2 \mathbf{x} +\int{ D \nabla q^* \cdot \nabla c}\,d^2\mathbf{x}  -\int{
	q^* \nabla \cdot\left(c\mathbf{v}  - \frac{D c}{k_B T}\nabla \phi\right)}\,d^2\mathbf{x} = -\int{\nabla q^* \cdot \left( \sqrt{2 D c} \boldsymbol{\zeta}  \right)}d^2\mathbf{x}\,.   
 \label{eq:weakeq}
\end{equation}
We discretize the solution and the test function in space using a set of linear shape functions based on Lagrange polynomials, which are defined on the linear quadrilateral elements of the computational mesh. The solution is approximated as
\begin{equation}
c\approx\sum_{i=1}^{N_{\text{dof}}}{c_i N_i(\mathbf{x})}\,,
\end{equation}
and the test function is approximated as
\begin{equation}
q\approx\sum_{i=1}^{N_{\text{dof}}}{q_i N_i(\mathbf{x})}\,,
\end{equation}
where the functions $N_i(\mathbf{x})$ are the shape functions and $N_{\text{dof}}$ represents the number of degrees of freedom. Replacing $c$ and $q$ in Equation \eqref{eq:weakeq}, we obtain a linear system of discrete equations that can be represented in matrix form as
\begin{equation}
\mathbf{L}\frac{\partial \mathbf{c}}{\partial t}  +(\mathbf{D}+\mathbf{A}(t))\mathbf{c}  = \mathbf{f}(t)\,,
\end{equation}
with 
\begin{equation}
{L}_{ij}=\int{ N_i N_j}d^2\mathbf{x}\,,
\end{equation}
\begin{equation}
{D}_{ij}=\int{D \nabla N_i\cdot \nabla N_j}d^2\mathbf{x}\,,
\end{equation}
and
\begin{equation}
{A}_{ij}(t)=\int{ N_i\cdot \nabla \left(\mathbf{v}(t) N_j-N_j \frac{D}{k_B T} \nabla \phi\right)}d^2\mathbf{x}\,,
\end{equation}
where the advection matrix $\mathbf{A}$ is a function of time since it depends on the advection velocity $\mathbf{v}$, which in turn depends on the time-dependent angular velocity $\omega$. The linear system is completed by enforcing the value at the degrees of freedom on the outer boundary, where Dirichlet boundary conditions $c=c_{\infty}$ are prescribed, and a zero flux on the boundary of the Janus disk.

To integrate in time, we use a semi-implicit Mayurama scheme. The semi-implicit time integration scheme has been shown to yield the correct structure factor for the fluctuations \cite{donev2010accuracy, martinez2024finite}. The time-integrated system can be written as
\begin{equation}
\left(\mathbf{L}+\frac{\Delta t}{2}(\mathbf{D}+\mathbf{A}(t_{n-1}))\right)\mathbf{c}(t_n)=  \left(\mathbf{L}-\frac{\Delta t}{2}(\mathbf{D}+\mathbf{A}(t_{n-1}))\right)\mathbf{c}(t_{n-1})+ \mathbf{f}(t_{n-1})\,,
\end{equation}
with the stochastic forcing term defined as
\begin{equation}
{f}_{i}(t_n)=-\sqrt{2 D \Delta t} \sum_{
k=1}^{N_g}{\sqrt{c(\mathbf{x}_k) w_k}\nabla N_i(\mathbf{x}_k)\cdot\mathbf{z}(\mathbf{x}_k,t_n)}\,,
\end{equation}
where $\Delta t$ is the time step, $N_g$ is the number of spatial integration points corresponding to degree of freedom $i$, $w_k$ is the integration weight corresponding to point $k$ and $\mathbf{z}$ is a vector with elements composed of Gaussian-distributed noise decorrelated in time and space, with zero mean and co-variance
\begin{equation}
  \left<{z}_l (\mathbf{x}_j,t_n){z}_m(\mathbf{x}_k,t_p)\right>=\delta_{lm} \delta_{jk} \delta_{np}  \,.
\end{equation}
The advection matrix needs to be updated for every time $t_n$, since it depends on the angular speed $\omega$, defined by Equation \eqref{eq:torquebal}, which can vary in time. The angular speed is computed for every time step as
\begin{equation}
\boldsymbol{\omega}(t_n)= - \sum_{i=1}^{N_{\text{dof}}}c_i(t_n) \int_{\Omega_f}{\mathbf{r} \times \frac{N_i  \nabla\phi}{\xi}} d^2\mathbf{x} \,.
\label{eq:omega3}
\end{equation}

\end{document}